\documentclass[PRX,reprint,amsmath,amssymb,superscriptaddress,longbibliography]{revtex4-1}

\usepackage{graphicx}
\usepackage{dcolumn}
\usepackage{bm}
\usepackage{hyperref}
\usepackage{breakurl}
\usepackage{multirow}
\usepackage{enumitem}

\begin{document}

\title{Negative Magnetoresistance in Weyl semimetals NbAs and NbP: Intrinsic Chiral Anomaly and Extrinsic Effects}

\author{Yupeng Li}
      \affiliation{Department of Physics, Zhejiang University, Hangzhou 310027, P. R. China}

\author{Zhen Wang}
      \affiliation{Department of Physics, Zhejiang University, Hangzhou 310027, P. R. China}
      \affiliation{State Key Lab of Silicon Materials, Zhejiang University, Hangzhou 310027, P. R. China}

\author{Pengshan Li}
      \affiliation{Institute of High Energy Physics, Chinese Academy of Sciences, Beijing 100049, P. R. China}

\author{Xiaojun Yang}
      \affiliation{Department of Physics, Zhejiang University, Hangzhou 310027, P. R. China}
      \affiliation{State Key Lab of Silicon Materials, Zhejiang University, Hangzhou 310027, P. R. China}

\author{Zhixuan Shen}
      \affiliation{Department of Physics, Zhejiang University, Hangzhou 310027, P. R. China}

\author{Feng Sheng}
      \affiliation{Department of Physics, Zhejiang University, Hangzhou 310027, P. R. China}

\author{Xiaodong Li}
      \affiliation{Institute of High Energy Physics, Chinese Academy of Sciences, Beijing 100049, P. R. China}

\author{Yunhao Lu}
      \affiliation{State Key Lab of Silicon Materials, Zhejiang University, Hangzhou 310027, P. R. China}

\author{Yi Zheng}
      \email{phyzhengyi@zju.edu.cn}
      \affiliation{Department of Physics, Zhejiang University, Hangzhou 310027, P. R. China}
      \affiliation{Zhejiang California International NanoSystems Institute, Zhejiang University, Hangzhou 310058, P. R. China}
      \affiliation{Collaborative Innovation Centre of Advanced Microstructures, Nanjing 210093, P. R. China}

\author{Zhu-An Xu}
      \email{zhuan@zju.edu.cn}
      \affiliation{Department of Physics, Zhejiang University, Hangzhou 310027, P. R. China}
      \affiliation{State Key Lab of Silicon Materials, Zhejiang University, Hangzhou 310027, P. R. China}
      \affiliation{Zhejiang California International NanoSystems Institute, Zhejiang University, Hangzhou 310058, P. R. China}
      \affiliation{Collaborative Innovation Centre of Advanced Microstructures, Nanjing 210093, P. R. China}

\date{\today}

\begin{abstract}
Chiral anomaly induced negative magnetoresistance (NMR) has been widely used as a critical transport evidence on the existence of Weyl fermions in topological semimetals. In this mini review, we discuss the general observation of the NMR phenomena in non-centrosymmetric NbP and NbAs. We show that NMR can be contributed by intrinsic chiral anomaly of Weyl fermions and/or extrinsic effects, such as superimposition of Hall signals, field-dependent inhomogeneous current flow in the bulk, i.e. current jetting, and weak localization (WL) of coexistent trivial carriers. Such WL controlled NMR is heavily dependent on sample quality, and is characterized by pronounced crossover from positive to negative MR growth at elevated temperatures, as a result of the competition between the phase coherence time and the spin-orbital scattering constant of the bulk trivial pockets. Thus, the correlation of NMR and chiral anomaly needs to be scrutinized, without the support of other complimentary techniques. Due to the lifting of spin degeneracy, the spin orientations of Weyl fermions are either parallel or antiparallel to the momentum, a unique physical property known as \textit{helicity}. The conservation of helicity provides strong protection for the transport of Weyl fermions, which can only be effectively scattered by magnetic impurities. Chemical doping of magnetic and non-magnetic impurities are thus more convincing in probing the existence of Weyl fermions than the NMR method.

\end{abstract}

\maketitle

\section{Introduction}

Quasi-particle excitations of massless Dirac fermions in solids were first proposed in graphene by Wallace in 1947 \cite{WallacePR1947PRdiracfermion}, and the idea of spin non-degenerated Weyl fermions was published even earlier by H. Weyl in 1929 \cite{Weyl1929Elektron}. However, the experimental exploration of relativistic fermions in solids remained as a niche field, until the discovery of anomalous quantum Hall effects of Dirac fermions in graphene \cite{GrapheGeim_QHE,GrapheKim_QHE}.  The birth of topological insulators \cite{HasanMZRMP2010TI,ZhangSCRMP2011TIandTS} has intrigued the intensive competitions in search of new topological semimetals (TSMs), such as Dirac semimetals (DSMs), Dirac nodal-line semimetals and Weyl semimetals (WSMs) \cite{DSM_PRLtheory_Kane,Cd3As2_PRBtheory_LMR,Cd3As2_NPOng_NatMat15,WSMWanXG_PRB,WSMDaiX_PRX,NCHasan_WSMTheory,PbTaSe2_Hasan_Nodal}. These materials not only host massless fermions, which are condensed-matter-physics realizations of the long-sought relativistic fermions in high-energy physics, but also show extraordinary physical properties for potential device applications. In general, three-dimensional (3D) DSMs with four-degenerate Dirac nodes near the Fermi level can lift the spin degeneracy of energy bands and transform into 3D Weyl semimetals, by breaking either time-reversal symmetry (TRS) or inversion symmetry (IS) \cite{HGB2012PRBtimereversaltoWeyl}. The latter has been theoretically predicted \cite{WSMDaiX_PRX,NCHasan_WSMTheory} and experimentally confirmed \cite{NbAsHasan_ARPES_NaturePhy,TaAsDingH_ARPESWSM} in the non-magnetic, IS-broken TaAs family, which has been considered by many as one major breakthrough in TSMs.

\begin{figure}[!thb]
\begin{center}
\includegraphics[width=3.5in]{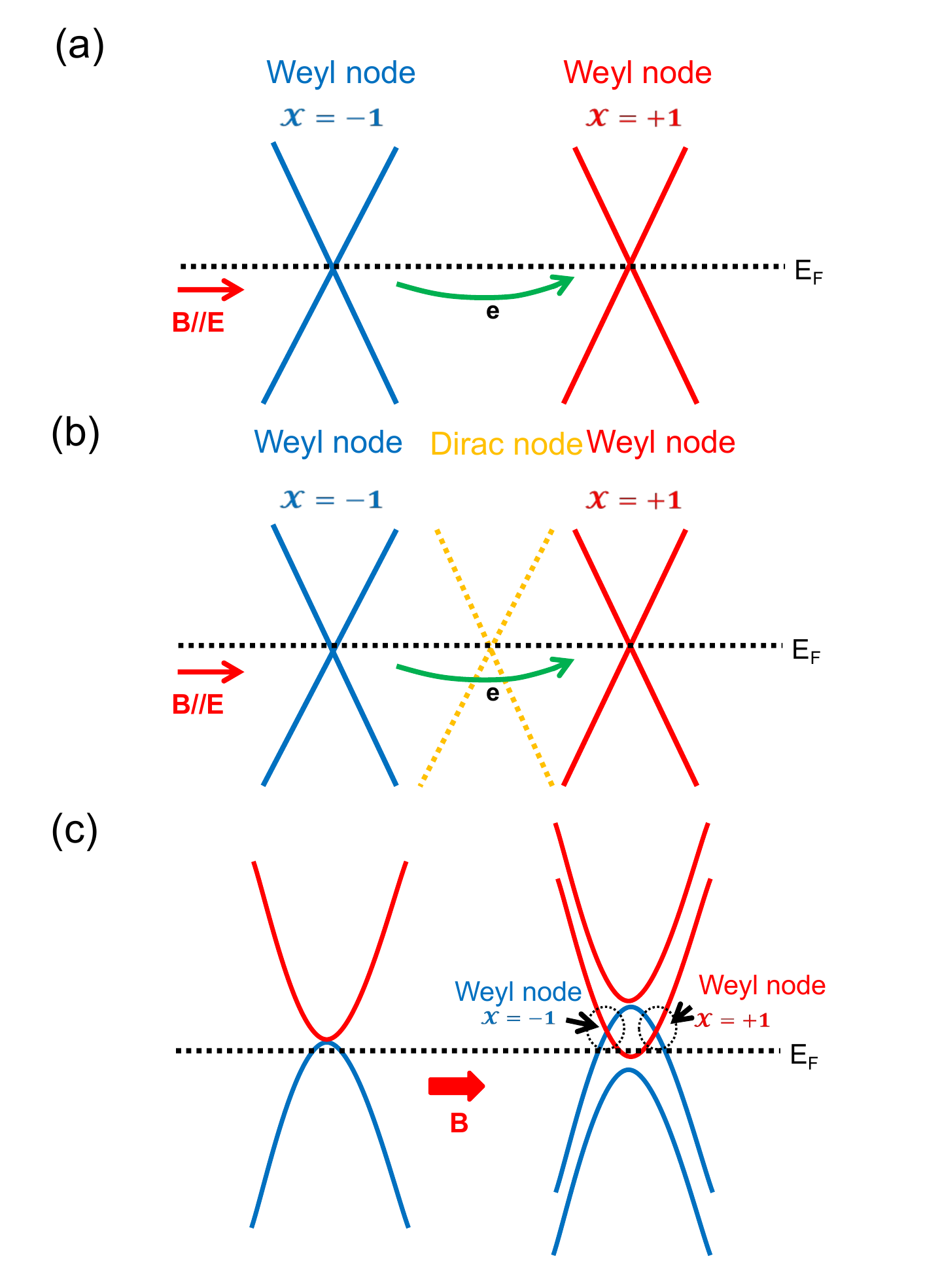}
\end{center}
\caption{\label{Fig1} Semiclassical chiral anomaly in WSMs, DSMs and half-Heusler semiconductors. (a) In WSMs, chiral anomaly emerges when $B$ is collinear with $E$, which creates extra conduction channels between two Weyl nodes with opposite chirality. (b) For DSMs, a Dirac cone may split into one pair of Weyl cones in magnetic field, which would explain the observed NMR phenomena. (c) Zeeman coupling may also lead to the formation of Weyl nodes in half-Heusler GdPtBi with accidentally touched quadratic bands, each becomes two spin non-degenerate bands in field.}
\end{figure}

Unlike the magnetic counterpart of pyrochlore iridates \cite{WSMWanXG_PRB}, WSM states in the TaAs family can be directly probed by angle-resolved photoemission spectroscopy (ARPES), as evident by the existence of Weyl node pairs and the linearly dispersed WSM bands \cite{TaAsDingH_ARPESWSM,TaAsDingH_ARPESNode,TaAsHasan_ARPES_Science,NbAsHasan_ARPES_NaturePhy}. WSMs also host symmetry protected topological surface states. The bulk Weyl nodes with opposite chirality of $\chi=+1$ and $\chi=-1$ are the source and drain points of Berry flux in momentum space, respectively. The projection of these paired singularities on any surface must be connected by open Fermi arcs \cite{WSMWanXG_PRB,WSMDaiX_PRX}, which have also been visualized by ARPES. Transport signatures of WSM states \cite{TaAs_arXiv_Jia,TaAs_PRX_NMR,NbP_NaturePhy,NbP_PRBWZ} have been reported almost simultaneously with the spectroscopy results. However, WSMs share common features with Dirac semimetals \cite{Cd3As2_NPOng_NatMat15,Cd3As2_PRLXMR_Coldea} in the transport measurements \cite{XLQi13_Physique_WSM}. For example, NbP shows one of the highest records of extremely large magnetoresistance (XMR), which is quasi linear and non-saturating in the field dependence \cite{NbP_NaturePhy,NbP_PRBWZ}. Similar XMR has also been reported in Dirac semimetal of Cd$_{3}$As$_{2}$ \cite{Cd3As2_NPOng_NatMat15,Cd3As2_PRLXMR_Coldea}. By measuring Shubnikov-de Haas (SdH) oscillations, a non-trivial Berry's phase of $\pi$ is expected for Weyl fermions, but such quantum geometrical phase is general for all quasiparticles associated to the massless linear spectrum \cite{3DBerryPhase_PRL04}. In the ultra quantum regime of strong magnetic field and low temperature, which was first considered by Nielsen and Ninomiya, two Weyl nodes of opposite chirality can exchange particles when the Fermi level lies within the zeroth Landau level. Such effect, well known as the Adler-Bell-Jackiw anomaly or chiral anomaly \cite{ChiralAnomaly_PLB}, manifests as negative magnetoresistance (NMR) when the external filed is collinear with the applied electric filed ($B\|E$), as illustrated in Figure \ref{Fig1}a.

However, anomalous NMR phenomena have been widely observed by transport measurements in DSMs, such as Cd$_3$As$_2$\cite{Cd3As2_NPOng_NatMat15}, Na$_3$Bi\cite{NPOngNMRNa3Bi}, Bi$_{1-x}$Sb$_x$\cite{Kim2013PRLAdlerBellJackiwAnomalyinTI}, and ZrTe$_5$\cite{LiQ2014NMRinZrTe5}, along with the transport reports of chiral anomaly in the TaAs-family WSMs \cite{TaAs_PRX_NMR,YangbhTaP15,YangxjNbAs15,NbP_PRBWZ}. This raises the question whether such NMR is from intrinsic chiral anomaly or due to extrinsic factors. One attempt to reconcile the discrepancy between the theoretical prediction and experimental observations is made by Kim \textit{et al}, who proposed that DSMs may transform into WSMs due to the lifting of spin degeneracy by the external magnetic field \cite{Kim2013PRLAdlerBellJackiwAnomalyinTI}, as illustrated in Fig. \ref{Fig1}b. Similar idea has also been adopted by Hirschberger \textit{et al} to explain the possible chiral transport in half-Heusler GdPtBi, which is zero-gap semiconductor with quadratic bands accidentally touching at the $\Gamma$ point \cite{NPOng_arXivGdPtBi}.

In real systems, the intrinsic Fermi levels of the TaAs-family WSMs are not located exactly at the energy of the chiral nodes. The resulting WSM Fermi surfaces are ellipsoids enclosing the chiral nodes. In such semiclassical regime, NMR can also arise from the non-zero Berry's curvature, which is satisfied by both DSMs and WSMs \cite{Spivak_ChiralAnomaly,Burkov_15PRB_NMR,Spivak_NMRModeling}. The resulting NMR is expected to be quadratic in the weak-field limit \cite{Burkov_15PRB_NMR}. In contrast, the experimental observations of NMR in DSMs and WSMs often extend to the high-field regime, and non-saturating behavior is reported by different groups \cite{YangbhTaP15,NPOng_arXivGdPtBi}. Such unusual field dependence, which can surprisingly persist above 100 K, strongly suggests that the current flow in these measured samples are inhomogeneous in the collinear configuration of $B$ and $E$, i.e. current jetting \cite{BettsJB2005PRLcurrentjet,RosenbaumTF2007PRBinhomogeneousconductors}. Using spot welding to make point-like contacts, dos Reis et al showed that current jetting plays a predominant role in the NMR of the TaAs family \cite{Hassinger_arXivNMRJetting}. The authors claimed that with $B\|E$, the field-dependent resistivity anisotropy $\rho_{zz}/\rho_{xx}$, which is the ratio of $\rho$ perpendicular and parallel to the field direction, can explain all the NMR characteristics reported in literatures \cite{Hassinger_arXivNMRJetting}. Contradictory, Zhang et al suggested that such field-dependent inhomogeneous current distribution can be effectively suppressed by using long and thin bar-shaped samples with electrodes fully crossing the width, thus the intrinsic chiral anomaly-induced NMR can be probed in the low field regime \cite{JiaShuang_NCommNMR}. However, analytical modelling of intrinsic chiral anomaly NMR becomes formidable when the specific defect structures of individual samples, which are dominantly high-density stacking faults in TaAs but a mixture of stacking faults, vacancies and anti-sites in TaP \cite{Besara_arXivTEM}, need to be taken into account.  

In this mini review, we overview our recent studies on the anomalous MR in NbAs and NbP, with an emphasis on the physical origins of the NMR phenomena with $B$ parallel to $E$. We elucidate the critical role of weak-antilocalization (WAL) and weak-localization (WL) of bulk trivial carriers in NbAs and NbP, which dominates the anomalous MR in low quality crystals and at elevated temperatures above 50 K. Unlike the massless Weyl fermions, which always contributes steep positive MR in low fields when a Berry's phase of $\pi$ is accumulated for time reserved scattering paths, the quantum correction of the trivial pockets is not only field- but also temperature-dependent. Consequently, the trivial pocket controlled MR is characterized by pronounced crossover behavior from low-field positive growth to NMR, as a result of the competition between WAL and WL of the trivial pockets. Surprisingly, such crossover behavior is also distinctive above 50 K in high-quality NbP, which shows order of magnitude higher charge carrier mobility than the other TaAs family members. This suggests the importance of understanding defects structures in these binaries, as TEM has revealed in TaAs and TaP recently. Systematic understanding of NMR and chiral anomaly will require the preparation of high-quality thin films of the TaAs family, to completely eliminate inhomogeneous current distribution and to achieve gate-tunable chiral anomaly. For transport experiments, a comparative chemical doping study of magnetic and non-magnetic impurities provides a more reliable way in probing the existence of WSM states. The helicity protection of Weyl fermions in IS-broken WSM can only be invalided by magnetic impurities, which has been demonstrated in our recent published results.

\section{Discussion}
\subsection{General NMR Characteristics: XMR Effect and Weak Antilocalization}

\begin{figure}[!thb]
\begin{center}
\includegraphics[width=3.5in]{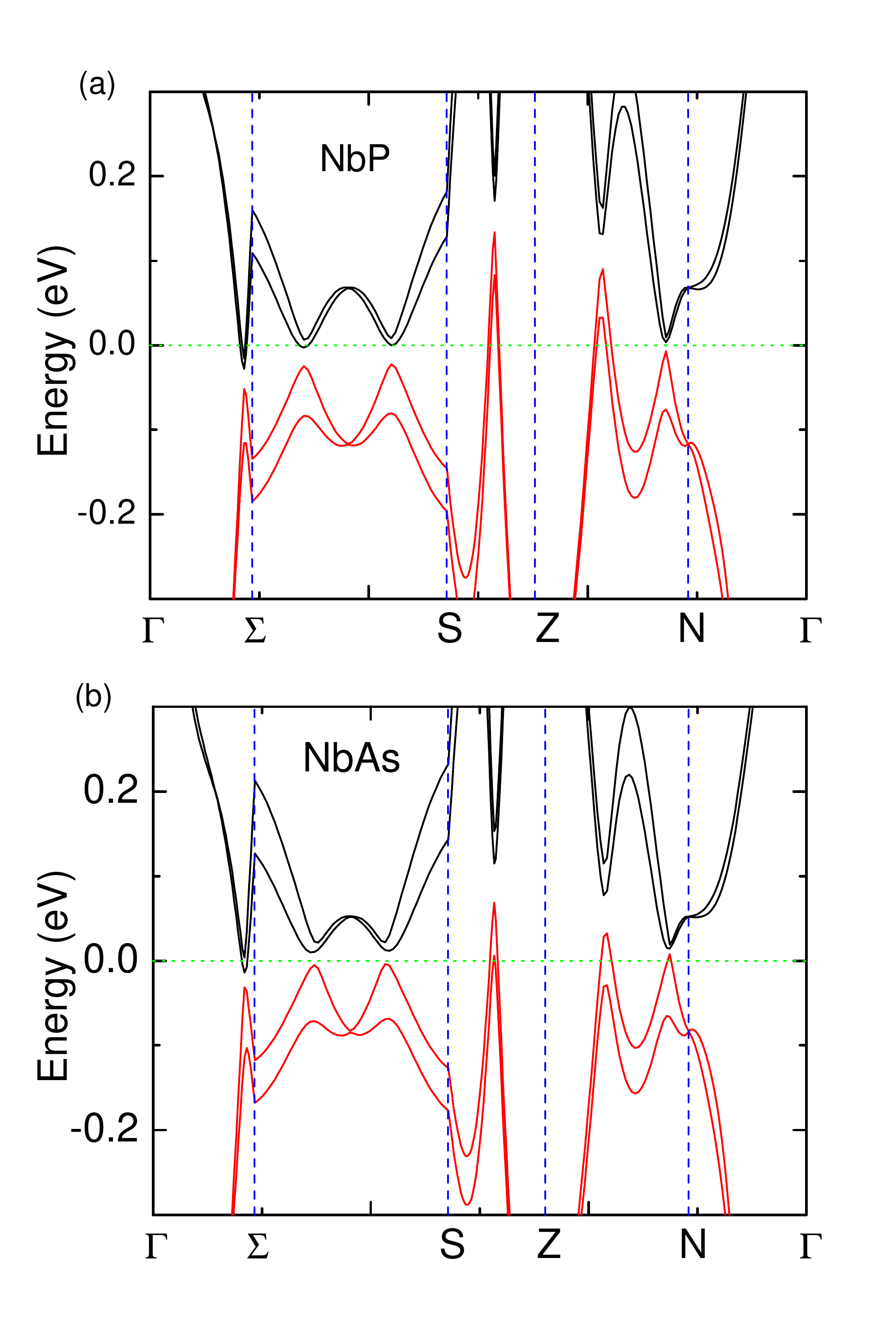}
\end{center}
\caption{\label{Fig2} The energy band structures of NbP (a) and NbAs (b). The change from P-3p to As-4p orbitals effectively reduces the trivial hole-pocket size and the overall charge carrier density.}
\end{figure}

We first briefly discuss the band structures of NbP and NbAs, which are noticeably different from the prototype TaAs. In NbP, there exist eight large trivial hole pockets, forming four pairs of inner and outer Fermi surfaces along the $S-Z$ symmetry line \cite{NbP_NaturePhy,NbP_PRBWZ}. Such trivial hole population is compensated by four n-type Weyl fermion pockets in the $k_{z}$ = 0 plane near the high symmetry points $\Sigma$, with each such WSM FS enclosing a trivial electron pocket \cite{NbP_PRBWZ}. In NbAs, the band top of the trivial hole pockets is closer to the Fermi energy, which effectively reduces the charge carrier concentration in the binary compared to NbP (Figure \ref{Fig2}). The change mainly comes from the hybridization of the As-4p and the Nb-4d orbitals, while the spin-orbital coupling (SOC) magnitude in both systems is comparable with the same transition-metal cations.

Due to the broken IS, the energy bands of NbP and NbAs are spin lifted. Thus, all charge carriers in these two compounds are spin polarized, similar to noncentrosymmetric WTe$_{2}$ as revealed by ARPES \cite{SOC_Aupes_PRL_FengDL}. For WSM pockets enclosing chiral nodes, the spin orientation is parallel or anti-parallel to the momentum, i.e. helicity, which is rooted in the massless Hamiltonian of WSMs, $\hat{H}=i\hbar v_{F} \overrightarrow{\sigma}\cdot \overrightarrow{K}$. Helicity provides extremely strong protection for the transport of Weyl fermions against the scattering of non-magnetic defects, and leads to spectacular charge carrier mobility of $1\times10^{7}$ cm $^{2}$V$^{-1}$s$^{-1}$ at 1.5 K in NbP with residual resistivity ratio (RRR) of  $\sim100$ \cite{NbP_PRBWZ}. For NbAs, we did not observe comparable mobility, which is about $3\times10^{5}$ cm $^{2}$V$^{-1}$s$^{-1}$ for high quality crystals with RRR of $\sim 100$. Distinctively, the anomalous MR of NbAs samples with $B\|E$ shows a steep MR upturn in low magnetic field, which is also generally observed in TaAs \cite{JiaShuang_NCommNMR} and TaP \cite{YangbhTaP15} at the helium temperature, but nearly indistinguishable in NbP \cite{NbP_PRBWZ}.

\begin{figure}[!thb]
\begin{center}
\includegraphics[width=3.5in]{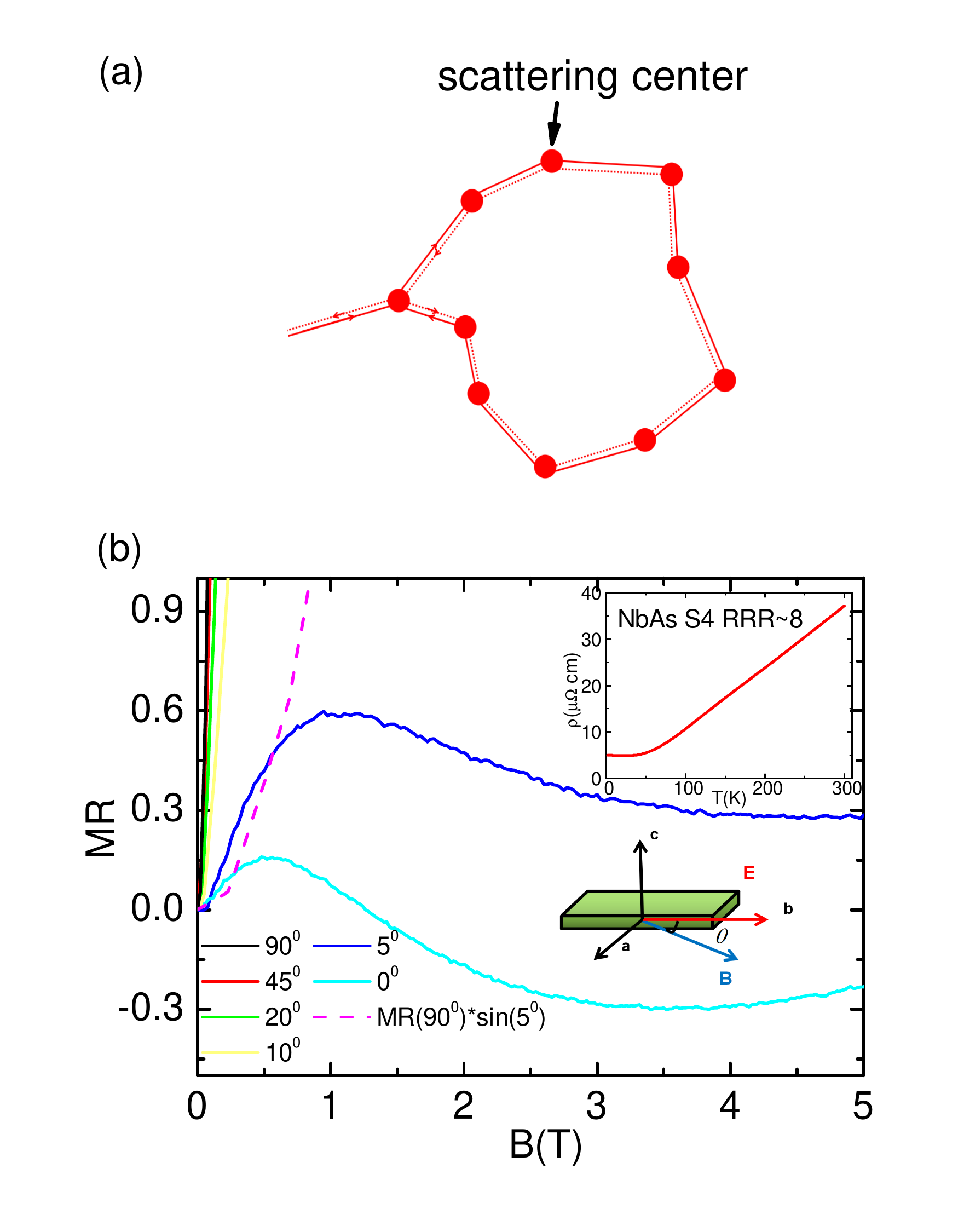}
\end{center}
\caption{\label{Fig3}  (a) Schematic of quantum interference induced by time reversed scattering paths.  (b) Angle-dependent anomalous MR in low-quality NbAs-S4. The pink dashed line is calculated from the transverse MR at 90$^{\circ}$, by assuming a large $B$ and $E$ misalignment of  5$^{\circ}$. Inset: Sample geometry: Single crystals of NbAs and NbP are polished into long and thin bar-shaped slabs with current contacts fully covering both long ends. The voltage leads are not point contacts on the top surface of the slab, but across the whole side surface along the $c$-axis to minimize the current jetting effect. Note that $E$ and $B$ are in the same plane during the angle rotation, and $a$- and $b$-axis are equivalent due to the lattice symmetry. }
\end{figure}

One possible contribution of such low-field MR hump is the XMR effect induced by a small misaligned angle between $B$ and $E$. In our experiments, we adopted a different angle rotation configuration from the literatures as shown in Figure \ref{Fig3}, in which $E$ is along the $b$-axis and $B$ is rotated in the $a-b$ plane. The zero degree is defined as the minimum point of the averaged MR vs angle curve for $B=\pm3$ T respectively, measured by a rotation step of $0.5^{\circ}$. As shown in Fig. \ref{Fig3}, at the presumptive $0^{\circ}$, the anomalous MR in NbAs is characterized by a low-field resistance hump, followed by the intermediate-field NMR growth. Above 3 T, the MR curve resumes slow positive growth, an indication of good angle alignment by our method. As a comparison, we plotted $\sin 5^{\circ}$ of the transverse MR ($90^{\circ}$, for $I\|b$ and $B\|a$) as the pink dashed line in Fig. \ref{Fig3}, which would be an extreme case of poor $B$ and $E$ alignment. It is clear that the low-field positive MR grows faster than the parabolic-like curve, which demonstrates that the low-field MR cannot be simply explained by the XMR effect.

\begin{figure}[!thb]
\begin{center}
\includegraphics[width=3.5in]{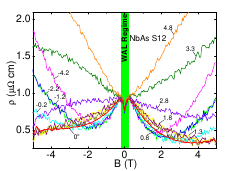}
\end{center}
\caption{\label{Fig4} Angle-dependent anomalous MR in high-quality NbAs-S12, in the vicinity of $B\|E$. The definition of the nominal $0^{\circ}$ is explained in the maintexts. Distinctively, the steep low-field MR hump is independent of angle rotations.}
\end{figure}

It is intriguing that the steep MR upturn is also sample quality dependent. In Figure \ref{Fig4}, we show the angle-dependent MR curves in the vicinity of $0^{\circ}$ for a high-quality sample with RRR of exceeding 130 (NbAs-S12), in contrast to RRR$\sim$8 for the sample in Fig. \ref{Fig3} (NbAs-S4). Despite that the overall MR characteristics are highly sensitive to the rotation angle ($\theta$), the low-field parts between $\pm0.2$ T are nearly identical. Above 0.2 T, the dwindling of the XMR contribution to the anomalous NMR in NbAs is evident when $\theta$ changes from $-4.2^{\circ}$ to $-0.2^{\circ}$. The nominal $0^{\circ}$ MR curve surprisingly shows weaker field-dependence in NMR growth, which deviates from 1/$B^{2}$ and is extended to much higher fileds compared the negative angles. Further increase in $\theta$ gradually weakens the field-dependence of NMR, which becomes non-saturating at $1.3^{\circ}$, $1.8^{\circ}$ and $2.8^{\circ}$. At angles lager than $3.3^{\circ}$, NMR is completely absent, and positive MR growth dominates above 0.2 T.

Although these experimental results could be interpreted in various ways as proposed in literatures \cite{JiaShuang_NCommNMR,YangbhTaP15,Hassinger_arXivNMRJetting}, the $\theta$-independent low-filed MR growth suggests that it is not dominated by the XMR effect or current jetting, both are sensitive to the $B$ and $E$ alignment. With the presence of massless Weyl fermions and spin-polarized trivial carriers, quantum inference induced MR correction would be expected. Such quantum effect is illustrated in Fig. \ref{Fig3}a, which shows a closed scattering trajectory and its time reversed path. For Weyl fermions, these two time reversed paths are equivalent to a circle movement in the momentum space, which leads to a Berry's phase of $\pi$ \cite{Bliokh_05PLAWAL}. Consequently, destructive quantum interference, i.e. weak antilocalization (WAL), is typical for WSM pockets at low temperatures. For trivial carriers, the conservation of spin-momentum locking, which is not well-defined helicity, is determined by the nature of individual scattering events. The resulting quantum correction is a pronounced competition between WAL and weak localization (WL), determined by the spin-orbital scattering time ($\tau_{SO}$) and the phase decoherence time ($\tau_{\phi}$), respectively. When non-magnetic impurity scattering is dominant, i.e. very short $\tau_{\phi}$, quantum interference is always in phase and constructive, producing WL correction to MR \cite{HLN_80ProgTheorPhys}. On the other hand, with $\tau_{\phi}\gg \tau_{SO}$, the quantum correction is antiphase and WAL dominates. In the intermediate regime of $\tau_{\phi}\sim \tau_{SO}$, there could be a field-dependent transition from WAL to WL \cite{WALWL_SciRep_WangJ14}.

\begin{figure}[!thb]
\begin{center}
\includegraphics[width=3.5in]{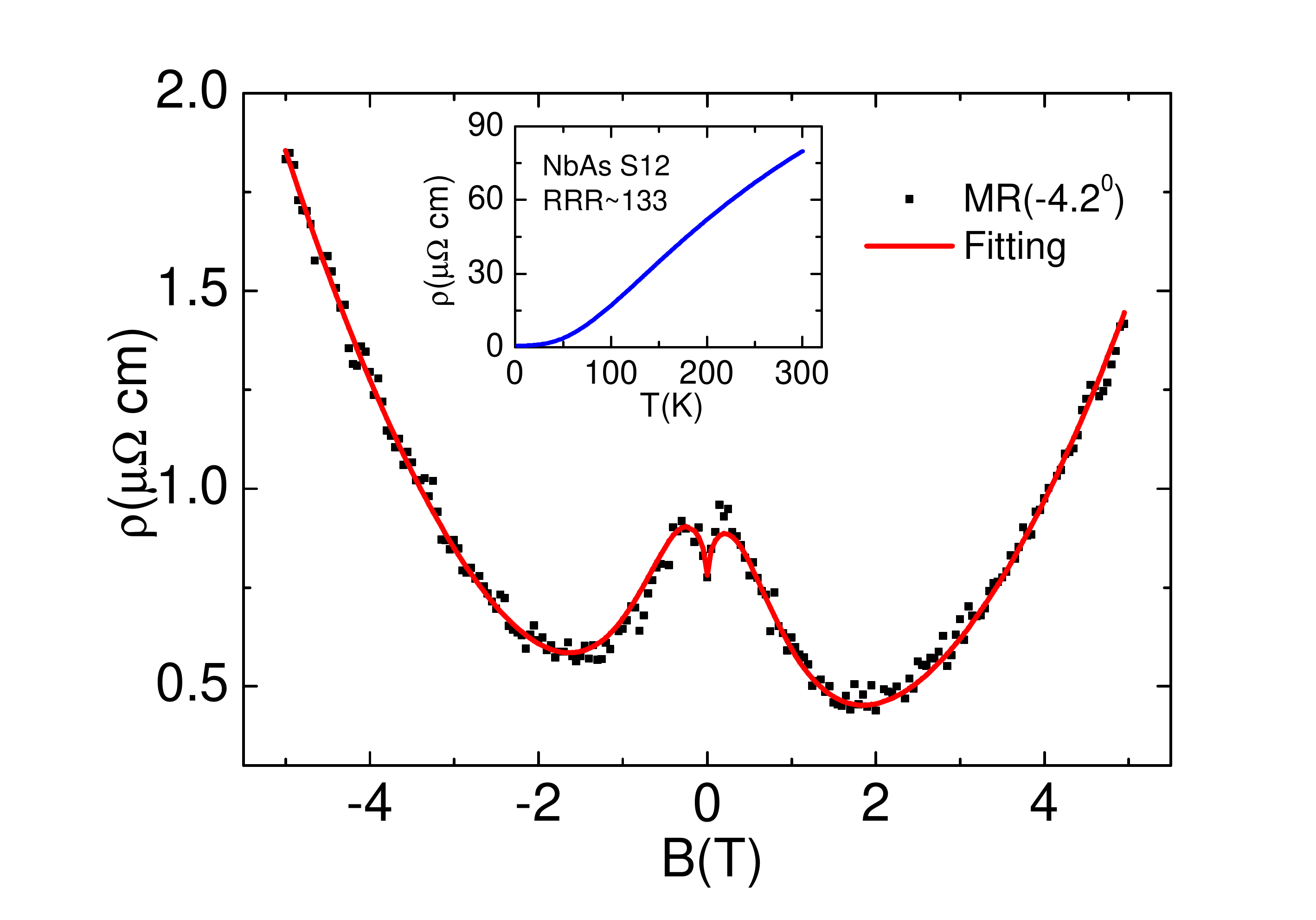}
\end{center}
\caption{\label{Fig5} Fitting of the anomalous MR curve at $-4.2^{\circ}$ for NbAs-S12, using the Eq. \ref{Eq2} in the main texts. The low-field MR follows the $a\sqrt{B}$ model, while the intermediate-field NMR growth is inversely proportional to $B^{2}$. Note that a linear Hall or current-jetting signal correction is superimposed on the anomalous MR characteristics.}
\end{figure}

In both $\tau_{\phi}\gg \tau_{SO}$ and $\tau_{SO}\gg \tau_{\phi}$ regimes, the quantum correction to conductivity with collinear $B$ and $E$ can be formulated by \cite{LiYQ13_PRB_WAL},
\begin{equation}
\label{Eq1}
\triangle \sigma(B)\simeq \alpha \frac{-e^{2}}{2\pi^{2}\hbar}ln(1+\beta\frac{ed^{2}}{\hbar B_{\phi}}B^{2}),
\end{equation}
in which $\hbar$ is the Planck constant, $e$ is the electron charge, $d$ is the sample thickness, $B_{\phi}$ is the critical field required to dephase the quantum interference, and $\beta$ is a fitting parameter. The difference between WAL and WL lies in the coefficient $\alpha$, which is $1/2$ for the former and $-1$ for the latter respectively. For practical purpose, the low-field WAL correction can be simplified as $\sigma_{WAL}=\sigma_{0}+a\sqrt{B}$ \cite{Kawabata80_JPSJ_WAL,Kim2013PRLAdlerBellJackiwAnomalyinTI}. By assuming a parabolic chiral magnetic conductivity of $C_{W}B^{2}$ \cite{Burkov_15PRB_NMR,Kim2013PRLAdlerBellJackiwAnomalyinTI}, we can fit the anomalous MR using an empirical model,
\begin{equation}
\label{Eq2}
\rho(B)=\frac{1}{\sigma_{WAL}\cdot (1+C_{W}B^{2})}+A_{1}B^{2}+A_{2}B,
\end{equation}
where $\sigma_{0}$ is the zero-field conductivity, $A_{1}B^{2}$ is the XMR contribution due to $\theta$ misalignment, and $A_{2}B$ is a linear correction to the resistivity, which may be due to the large Hall signals in NbAs or current jetting. The validity of Equation \ref{Eq2} is justified by the good alignment between $B$ and $E$, which makes sure that the XMR contribution is negligible in low fields. Indeed, this model captures all the essential features of the anomalous MR, including the low-field MR hump below 0.2 T, followed by the parabolic NMR, and positive MR growth above 2 T, as shown in Figure \ref{Fig5} for the $-4.2^{\circ}$ data in Fig. \ref{Fig4}. However, for $0^{\circ}$ and positive $\theta$, Eq. \ref{Eq2} does not fit the NMR growth, which deviates from 1/$C_{W}B^{2}$ significantly.

\begin{figure}[!thb]
\begin{center}
\includegraphics[width=3.5in]{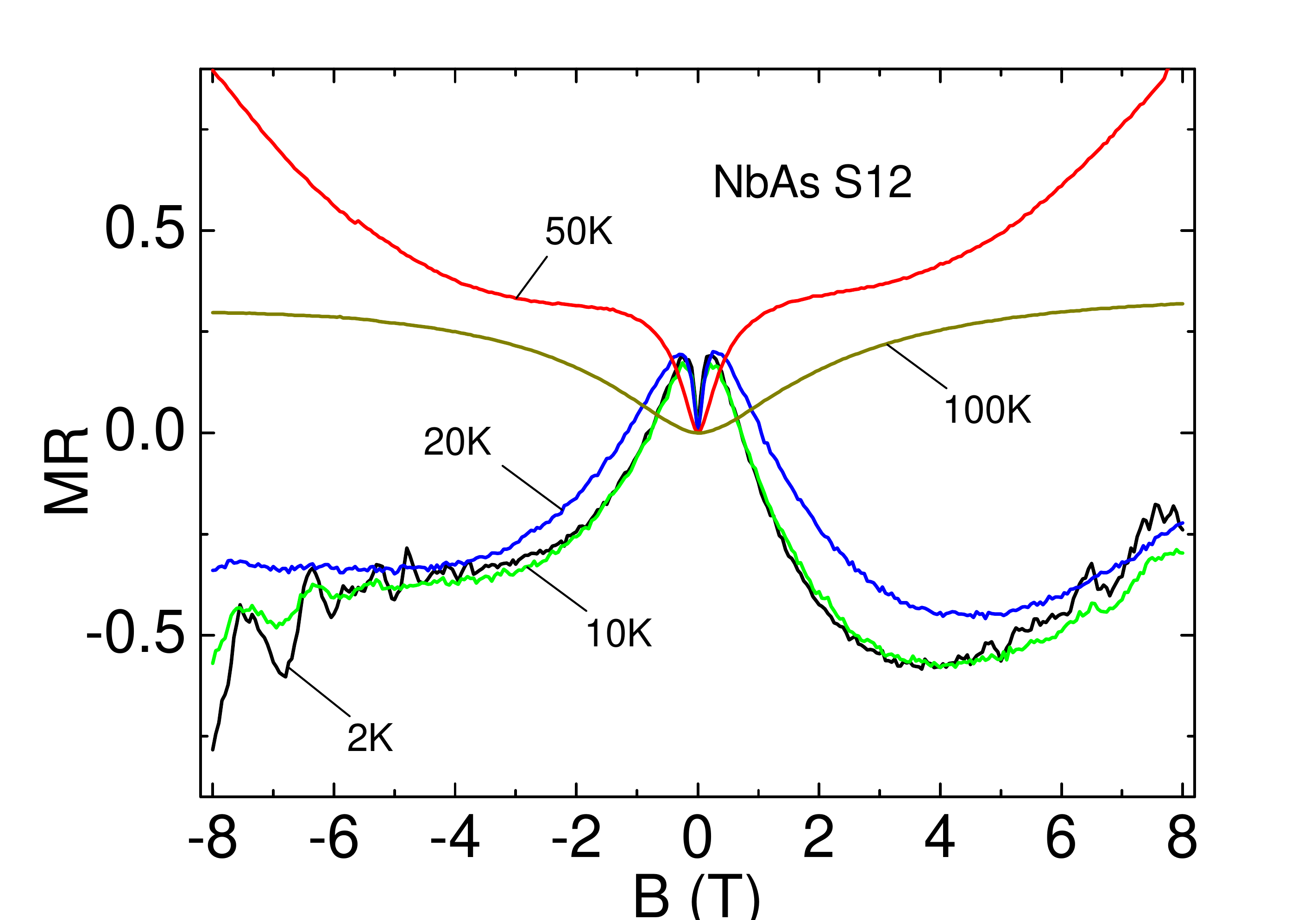}
\end{center}
\caption{\label{Fig6} $T$-dependent WAL in high-quality NbAs-S12. At 50 K, the MR curve clearly shows the WAL-related positive growth, followed by the parabolic-like XMR effect. Note that NMR is completely suppressed at this temperature. At 100 K, typical XMR behavior is observed, which becomes saturating in high fields.}
\end{figure}

For WSM associated WAL, it is also expected to remain robust at elevated temperature, although electron-phonon, and intravalley electron-electron interactions will weaken the helicity protection mechanism.  In Figure \ref{Fig6}, we show the T-dependent MR at the nominal $0^{\circ}$ for NbAs-S12. It is clear that NMR has stronger $T$ dependence than WAL, which is nearly identical below 20 K. At 50 K, NMR disappears completely, while low-field WAL, which becomes broader in field dependence, is followed by parabolic-like XMR. Above 50 K, MR is dominated by the XMR effect, which becomes saturating in high field like the conventional semimetals \cite{Kopelevich03_PRL_GraphiteMR,Behnia09_BismuthMR}.

\subsection{Non-saturating NMR: Contribution of Weak Localization from Trivial Pockets}
\begin{figure}[!thb]
\begin{center}
\includegraphics[width=3.5in]{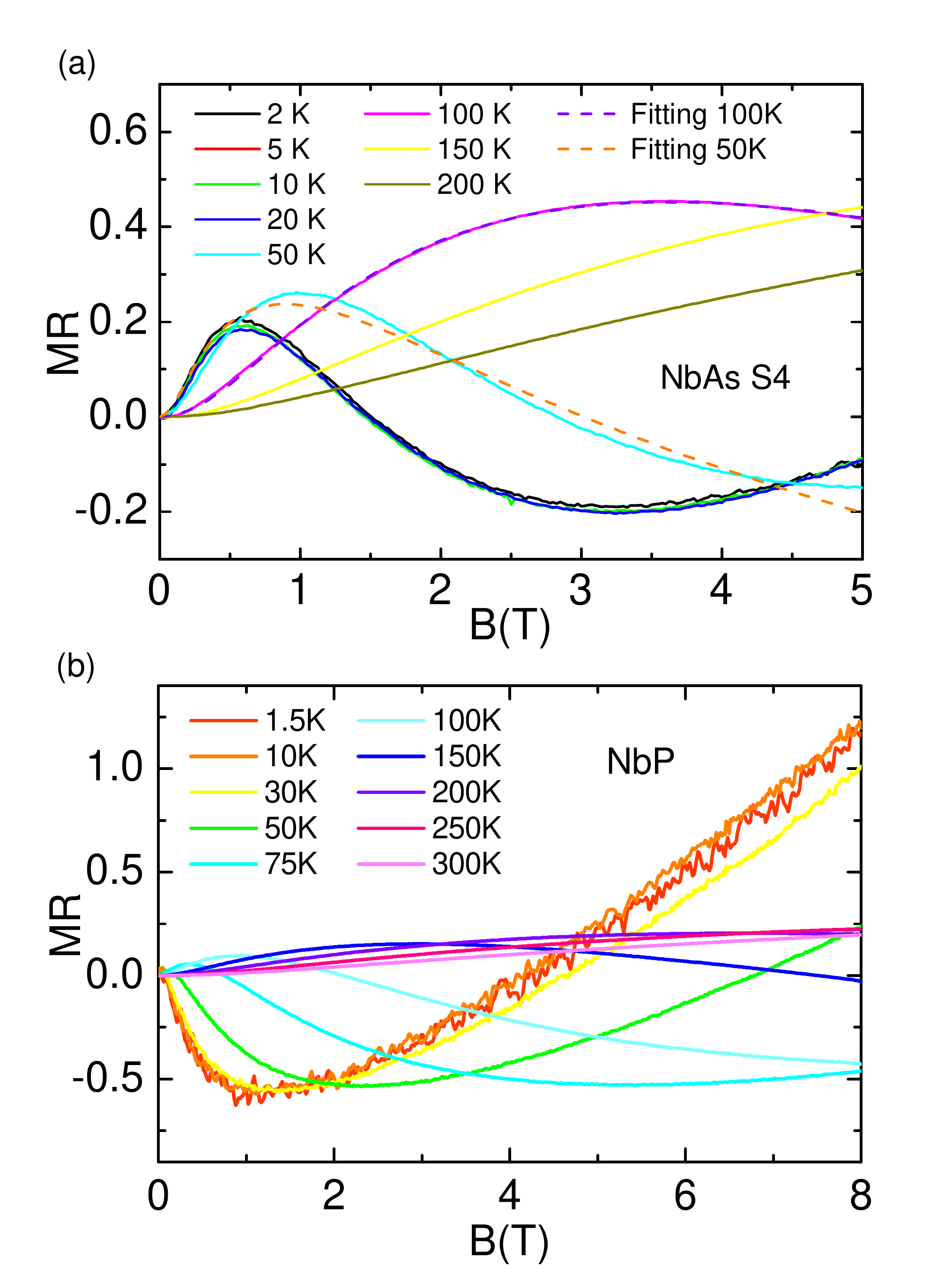}
\end{center}
\caption{\label{Fig7} (a) $T$-dependent anomalous MR in low-quality NbAs-S4. The low-field positive MR is significantly different from the high-quality sample in Fig. \ref{Fig6}. (b) $T$-dependent anomalous MR in high-quality,  stoichiometric NbP. At helium temperature, the low-field positive MR cusp is nearly indistinguishable.  Above 50 K, broad WAL-WL crossover characteristics are dominant.}
\end{figure}

Noticeably, sample quality plays a critical role in the $T$-dependent anomalous MR in NbAs. In Figure \ref{Fig7}, we show the $T$-dependent MR curves for NbAs-S4, which has one order of magnitude lower RRR of $\sim8$ compared to NbAs-S12. The low-field positive MR cusp in this sample is largely extended to 0.5 T and shows significant $T$ dependence below 20 K. Unlike the high-quality NbAs-S12, there are broad transitions from positive to negative MR in NbAs-S4. At 50 K, NMR in NbAs-S4 becomes non-saturating up to 5 T. Further increase in $T$ to 100 K, the positive MR growth becomes parabolic-like, suggesting its origin in the XMR effect. However, above 3 T, the negative MR growth gradually takes over. Li et al proposed that such unusual crossover feature is correlated to pronounced field-dependent competition between the WAL and WL of trivial bulk pockets \cite{YupengLiNbAs22016NMR},  as reported in the MPn$_2$ family (M = Nb and Ta; Pn = As and Sb) \cite{YupengLiNbAs22016NMR,ShengbingNbAs22016NMR,LuoykTaAs22016,WangzTaSb22016}. Due to the coexistence of non-trivial and trivial pockets, such mechanism would be general for topological SMs with strong SOC.

In the regime of $\tau_{\phi}\sim \tau_{SO}$ \cite{WALWL_SciRep_WangJ14}, such trivial pocket contributed broad crossover with $B\|E$ can be modelled by assuming two independent scattering processes for WAL and WL,
\begin{equation}\label{DKmodel}
\frac{\triangle R}{R_{0}}=\alpha[\frac{1}{2}\ln(1+bB^{2}L_{\phi}^2)-\frac{3}{2}\ln(1+bB^{2}L_{SO}^2)],
\end{equation}
in which $L_{SO}\propto \tau_{SO}$ and $L_{\phi} \propto \tau_{\phi}$ are the spin-orbital scattering length and the phase coherent length respectively. WAL is dominant in low fields when $\tau_{\phi}>\tau_{SO}$. However, WL gradually takes over in high fields due to a larger coefficient of $3/2$ than $1/2$ for WAL. More importantly, $\tau_{SO}$ is an intrinsic parameter of SOC in the bulk, and thus is weakly dependent on temperature changes, while $\tau_{\phi}$ is highly sensitive to $T$. Such distinct $T$ dependence between $\tau_{SO}$ and $\tau_{\phi}$ leads to pronounced $T$-dependent crossover characteristics. In Fig. \ref{Fig7}, we show the data fitting of 50 K and 100 K using Eq. \ref{DKmodel}. At 100 K, the model gives a satisfying description on the field dependence. In contrast, the fitting of the 50 K result is less successful, which implies the contribution of chiral anomaly or other extrinsic effects to NMR.

\begin{figure}[!thb]
\begin{center}
\includegraphics[width=3.5in]{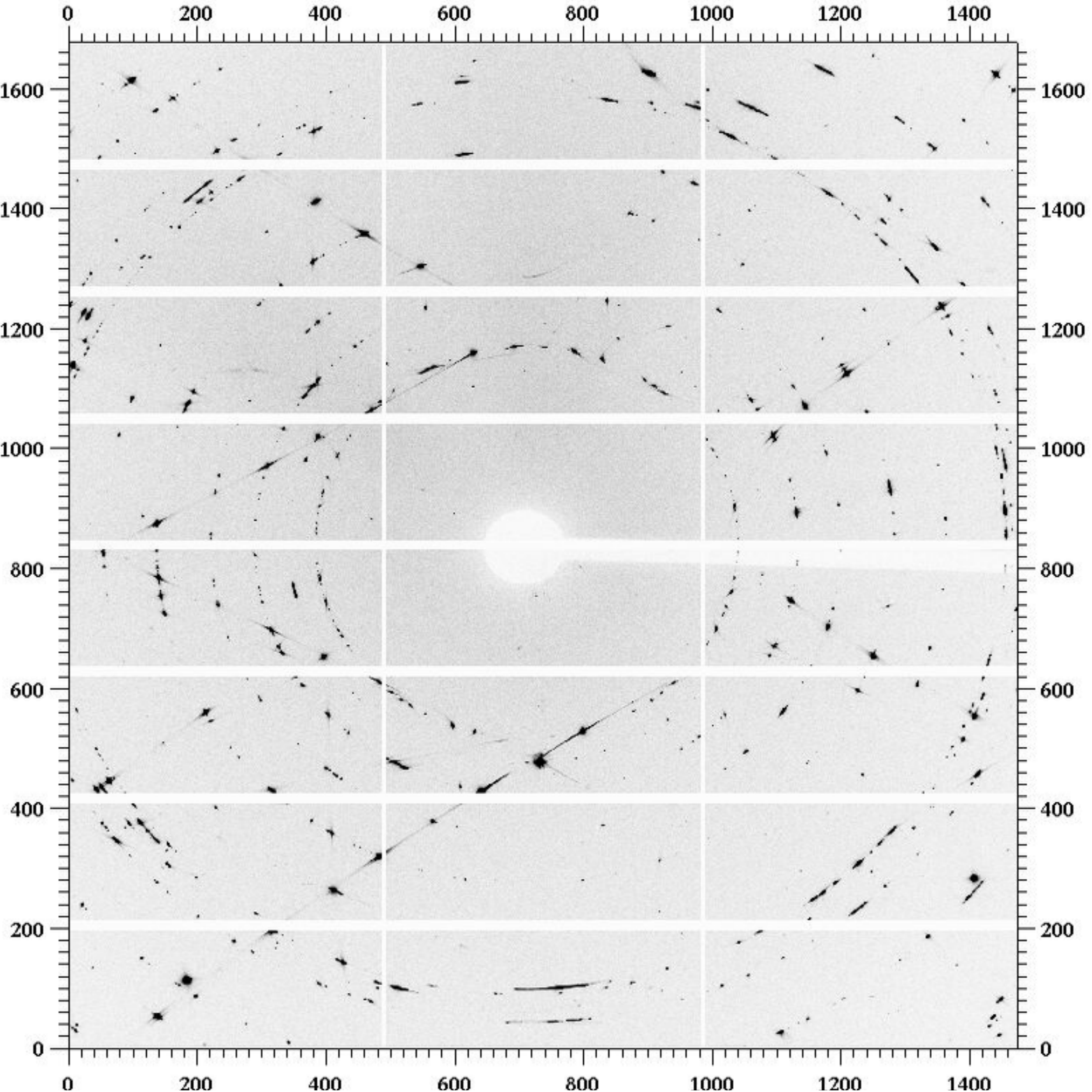}
\end{center}
\caption{\label{Fig8} Angle-dispersive x-ray diffraction of a low quality NbAs single crystal, showing typical stacking fault lines. Similar defects have also been reported in TaP and TaAs by scanning transmission electron microscopy \cite{Besara_arXivTEM} .}
\end{figure}

It should be emphasized here that Eq. \ref{DKmodel} is a variation of the Dugaev-Khmelnitskii model \cite{DKmodel_JETP84}, which in principle requires $d\ll l$. Here $l$ is the mean free path of trivial carriers, which is several orders of magnitude smaller than the typical sample thickness of 200 $\mu m$ in our experiments. The applicability of Eq. \ref{DKmodel} to thick NbAs samples may be correlated to the formation of unique stacking-fault defects in the $a-b$ plane, which essentially produces quasi-2D structure of $I4_{1}md$ lattice sandwiched by hexagonal lattice of $P\bar{6}m2$ \cite{Besara_arXivTEM}. In Figure \ref{Fig8}, we show the typical angle-dispersive x-ray diffraction pattens of a low-quality NbAs single crystal, which clearly reveals the stacking fault-related streaks. For NbP, highly stoichiometric single crystals can be synthesized, as manifested by the unprecedented charge carrier mobility of $1\times10^{7} \,cm^{2} V^{-1} s^{-1}$ \cite{NbP_PRBWZ}. However, we have also observed broad WAL-WL crossover in NbP above 50 K \cite{NbP_PRBWZ}, which implies that stacking-fault defects may be universal for the TaAs family.

\subsection{Probing Weyl Fermions by Magnetic Impurities}
\begin{figure}[!thb]
\begin{center}
\includegraphics[width=3.5in]{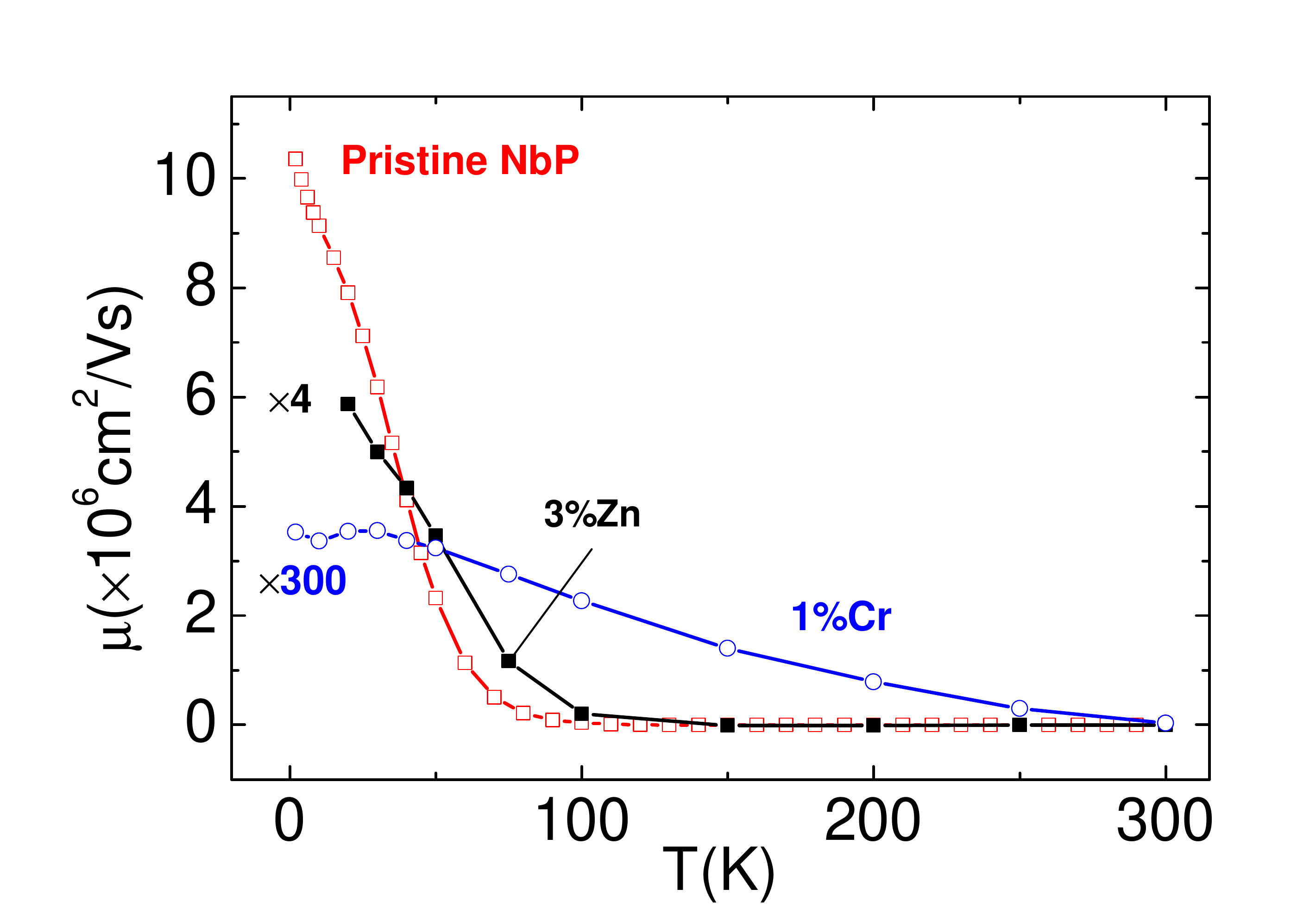}
\end{center}
\caption{\label{Fig9} Magnetic vs non-magnetic impurities in NbP. The ultrahigh charge carrier mobility of NbP is degraded by more than two orders of magnitude by $\sim 1\%$ Chromium, but insensitive to Zinc impurities.}
\end{figure}

Since the helicity protection of Weyl fermions in the TaAs class relies on intact time reversal symmetry, comparative chemical doping of non-magnetic and magnetic impurities in this family provides an indispensable way in probing the existence of Weyl fermions. In our previous study \cite{NbP_PRBWZ}, we have shown that a minimal amount of $\sim1\%$ Chromium in NbP will degrade charge carrier mobility by more than two orders of magnitude, while three times higher concentration of non-magnetic Zinc yields comparable mobility to pristine NbP (See Figure \ref{Fig9}). Compared to the parallel-field NMR method, which may be completely dominated by multiple sources of extrinsic effects instead of chiral anomaly, chemical doping experiments would be an effective transport evidence on the existence of WSM states. Equally importantly, the interaction of magnetic impurities with Weyl fermions at low temperatures may lead to novel Kondo physics of massless fermions \cite{Fritz15_PRB_KondoWSM}, which has not been experimentally explored.

\section{Conclusion}
The discovery of WSM states in the IS-broken TaAs class has been a major breakthrough in condensed matter physics. The experimental observations of NMR have been extensively used as a fingerprint of WSM states in TaAs and the other three binaries. However, the recent experiments have raised the question of whether NMR can be used an unambiguous transport evidence in proving the existence of Weyl fermions. Here, we have discussed the general observation of anomalous MR in NbP and NbAs in the configuration of collinear $B$ and $E$, which is a prerequisite of chiral anomaly. We have elucidated that the low-field steep positive MR is WAL correlated, while the followed intermediate-field NMR may be contributed by intrinsic chiral anomaly and/or various extrinsic effects. In particular, we highlight WL contribution to NMR from trivial pockets in NbAs and NbP, coexisting with WSM states. The trivial pocket controlled WL is sensitive to crystal quality, and low RRR samples show more pronounced extrinsic NMR, but with weaker field dependence than the theoretical parabolic chiral magnetoconductance. At elevated temperatures, WL contributed NMR becomes predominant and shows non-saturating behavior in field growth, which would be general for topological SMs with strong SOC and coexisting non-trivial and trivial pockets.

To probe the existence of WSM states, chemical doping of non-magnetic and magnetic impurities provides a more reliable way by detecting the helicity protection mechanism of Weyl fermions. Note that such method is also valid for other relativistic fermions with spin-momentum locking, such as helical Dirac fermions in Rashiba SMs and TIs. A complete understanding of chiral anomaly induced NMR would be feasible by preparing the TaAs class into thin films, using methods such as chemical vapour deposition and pulse laser deposition. With thin film samples, it is also possible to introduce gate tunability to study Fermi energy dependent chiral anomaly in this family, which would allow the ultra quantum-limit chiral magnetoconductance to be studied for the first time.

\begin{acknowledgments}
This work was supported by the National Basic Research Program of China (Grant Nos. 2014CB921203), the National Science Foundation of China (Grant Nos. 11190023, U1332209, 11374009, 61574123 and 11574264), MOE of China (Grant No. 2015KF07), the Fundamental Research Funds for the Central Universities of China, and the National Key R\&D Program of the MOST of China (Grant No. 2016YFA0300204). The in-situ high-pressure angle-dispersive x-ray diffraction (ADXRD) measurement was performed at the 4W2 beamline of the Beijing Synchrotron Radiation Facility (BSRF). Y.Z. acknowledges the start funding support from the Thousand Talents Plan.
\end{acknowledgments}

\end{document}